\documentclass{emulateapj}
\def\degpoint{\ifmmode ^{\rm{o}}\!. \else $^{\rm{o}}\!.$\fi}
\newcommand{\degrees}{$^{\rm{o}}$}
\newcommand{\ms}{\mbox{m\ s$^{-1}$}}

\newcommand{\Msun}{\mbox{M$_{\odot}$}}

\newcommand{\Mjup}{\mbox{M$_{\rm Jup}$}}

\newcommand{\gtsimeq}{\raisebox{-0.6ex}{$\,\stackrel
         {\raisebox{-.2ex}{$\textstyle >$}}{\sim}\,$}}

\begin{document}
\title{Long-Period Objects in the Extrasolar Planetary Systems 47~UMa 
and 14~Her\footnotemark[1] 
}
\author{Robert A.~Wittenmyer, Michael Endl, William D.~Cochran}
\affil{McDonald Observatory, University of Texas at Austin, Austin, TX 
78712}
\email{
robw@astro.as.utexas.edu}

\shorttitle{Long-Period Companions}
\shortauthors{Wittenmyer, Endl, Cochran}
\slugcomment{Accepted for publication in ApJ}
\footnotetext[1]{Based on observations obtained with the Hobby-Eberly
Telescope, which is a joint project of the University of Texas at Austin,
the Pennsylvania State University, Stanford University,
Ludwig-Maximilians-Universit{\" a}t M{\" u}nchen, and
Georg-August-Universit{\" a}t G{\" o}ttingen.}

\begin{abstract}

\noindent The possible existence of additional long-period planetary-mass
objects in the extrasolar planetary systems 47~UMa and 14~Her is
investigated.  We combine all available radial-velocity data on these
stars, spanning up to 18 years.  For the 47~UMa system, we show that while
a second planet improves the fit to all available data, there is still
substantial ambiguity as to the orbital parameters of the proposed
planetary companion 47~UMa~c.  We also present new observations which
clearly support a long-period companion in the 14~Her system.  With a
period of 6906$\pm$70 days, 14~Her~c may be in a 4:1 resonance with the
inner planet.  We also present revised orbital solutions for 7 previously
known planets incorporating recent additional data obtained with the 2.7m
Harlan J.~Smith Telescope at McDonald Observatory.

\end{abstract}

\section{Introduction}

The longest-running radial-velocity surveys are now approaching time
baselines of 16-18 years \citep{butleretal96, cochran97}.  These surveys
now achieve internal measurement precisions such that the signals from
long-period giant planets ($P\gtsimeq 10$yr) are now entering the realm of
detectability \citep{limitspaper}.  For example, the High Resolution
Spectrograph (HRS) on the Hobby-Eberly Telescope (HET) now has a velocity
precision of 3-4 \ms \citep{cochran04}, the Keck HIRES is achieving 1-2
\ms\ since its 2004 CCD upgrade \citep{butler06}, and the HARPS instrument
has demonstrated precision better than 1 \ms \citep{lovis06}.  Of
particular interest are putative multi-planet systems, wherein the
residuals of a known-planet's orbit show Keplerian periodicity indicative
of a distant giant planet companion in the system.  Finding systems which
contain long-period giant planets as well as planets in closer orbits will
address important questions about the uniqueness of our own Solar System's
architecture.

Long-period planet candidates have been reported for 47~UMa ($P=7.1$ yr,
Fischer et al.~2002), 55~Cancri ($P=12.4$ yr, McArthur et al.~2004),
HD~217107 ($P>7$ yr, Vogt et al.~2005), and HD~72659 ($P=9.9$ yr, Butler
et al.~2006).  Of particular interest are systems which contain both
short-period and long-period jovian planets.  Such systems could provide
clues to address the question of how the processes of planet formation and
migration can result in both ``hot'' and ``cold'' Jupiters in the same 
planetary system.

In this paper, we present improved fits to the known planets in nine
systems using all available data, and we investigate the possibility of
additional long-period objects.  In \S 2, we briefly describe the data
acquisition and fitting procedure.  The results are given in \S 3,
including our solutions for seven additional planetary systems in the
McDonald Observatory planet search program.  All of our radial-velocity
measurements for these objects are presented in Tables 5-14.  In this
paper, we use the terms ``McDonald'' to refer to data taken with the
McDonald Observatory 2.7~m Harlan J.~Smith (HJS) Telescope, and ``HET'' to
refer to data taken with the 9.2~m Hobby-Eberly Telescope.

\section{Observations and Data Analysis}

Data obtained from the McDonald Observatory planet search program
\citep{endl05} are discussed fully in \citet{limitspaper}.  Available
published data were combined with the McDonald data to fit Keplerian
orbits using GaussFit \citep{jefferys87}.  GaussFit has the ability to
allow the offsets between data sets to be a free parameter.  Parameters of
the primary stars considered are given in Table 1; masses, [Fe/H], and
$T_{eff}$ are from \citet{santos04}, and the chromospheric activity index
log $R'_{HK}$ is derived from Ca~II measurements from the McDonald
Observatory spectra.  For each object, we searched for periodic signals in
the residuals to the known planet's orbit using the periodogram method
\citep{lomb76, scargle82}.  To assess the statistical significance of
those periods, the false alarm probabilities (FAP) were calculated using
the bootstrap randomization method detailed by \citet{kurster97} and
\citet{endl02}.

\section{Results and Discussion}

The Keplerian orbital fits are shown in Figures~1-4, and the orbital
parameters implied by those fits are given in Table 2.  In computing the
planetary minimum mass M~sin $i$ and semimajor axis $a$, the stellar
masses derived by \citet{santos04} listed in Table 1 were used, with
adopted uncertainties of 0.05 \Msun.

\subsection{47 UMa (=HD~95128): Ambiguities concerning the second planet}

\citet{butler96} first reported the 1090-day companion to 47~UMa using
data from Lick Observatory.  With additional velocity measurements over 13
years, \citet{fischer02} announced a long-period second planet, 47~UMa~c,
with a period of 2594$\pm$90 days and a mass of 0.76 \Mjup.  
\citet{naef04} presented ELODIE observations of 47~UMa, and noted that the
second planet was not evident in their data, which were fit well with a
single Keplerian model.

We now fit four data sets for 47~UMa: Lick (Fischer et al~2002, N=91),
ELODIE (Naef et al.~2004, N=44), 2.7m HJS telescope (N=35), and the
Hobby-Eberly Telescope (HET, N=201).  The HET data, which consist of
multiple exposures per visit, were binned using the weighted mean value of
the velocities in each visit.  The quadrature sum of the rms about the
mean and the mean internal error bar was adopted as the error bar of each
binned point (N=63).  The offset between the overlapping HJS and HET
datasets was used to merge them into one, which was used in all fits.  
The one-planet fit (Model~1) and the residuals to that fit are shown in
Figure~1.  We emphasize that this fit includes all available published
data, over a time span of more than 18 years, and includes 195
high-precision measurements obtained with the HET High-Resolution
Spectrograph at 61 independent epochs, which are given in Table 6.  The
total rms about the combined one-planet fit is 10.4 \ms.  The rms scatter
about the one-planet fit for each of the four datasets is:  Lick--10.7
\ms, ELODIE--13.0 \ms, HJS--13.5 \ms, HET--4.9 \ms.  A periodogram of the
residuals of all of the data to the 1-planet fit is shown in Figure 2.  
No clear peak rises above the noise level at any period between 2 and
10000 days; the total duration of the available data is now about 6900
days (18.3 years).  While a peak is present at about 2212 days, close to
the period reported for 47~UMa~c by \citet{fischer02}, its false-alarm
probability (FAP) is 27.2\%.

To further explore the possible presence of 47~UMa~c, we fit all of the
datasets with a two-planet model fixed at the parameters of
\citet{fischer02} (Model~2), and then repeated the fit allowing all
parameters to be free except for $e$ and $\omega$ of the second planet
(Model 3), which were fixed at 0.005 and 127\degrees, respectively, after
\citet{fischer02}.  No models achieved convergence with those two
parameters free.  The fit obtained by Model 3 is shown in the right panel
of Figure~2.  Model 2 had a reduced chi-square ($\chi^{2}_{\nu}$) of 5.81
and an rms of 12.7 \ms\ about the fit, whereas Model 3 had a
$\chi^{2}_{\nu}$ of 2.11 and an rms of 8.4 \ms.  For comparison, the
one-planet fit (Model 1) had a $\chi^{2}_{\nu}$ of 3.23 and an rms of 10.4
\ms.  Noting that the poor fit of Model 2 was largely due to errors in the
period of the inner planet, we re-did the fits allowing the parameters for
the inner planet to be free while fixing those of the outer planet at the
values reported by \citet{fischer02} (Model 4).  The $\chi^{2}_{\nu}$ of
this fit was 2.68, with an rms of 10.2 \ms.  These tests are summarized in
Table 3.  The free two-planet (Model 3) fit was the best of the four, in
terms of both the goodness-of-fit criterion ($\chi^{2}_{\nu}$) and the
total rms scatter about the fits.  The rms about the individual datasets
for this fit is the following: Lick--8.0 \ms, ELODIE--11.1 \ms, HJS--10.1
\ms, HET--5.3 \ms.  The parameters for the 47~UMa planetary system given
in Table~2 are those obtained by Model~3.  Although the best-fit set of
parameters obtained a period of 7586 days for the outer planet, there is
no corresponding peak on the periodogram shown in Figure~2.  However, the
periodogram method is not as reliable when the periodic signal approaches
or exceeds the total duration of observations, as is the case here, where
the total time baseline is 6942 days.  We note that we have been able to
reproduce the result of \citet{fischer02} by this method; i.e.~a
periodogram analysis of the Lick data alone after removing 47~UMa~b
revealed a strong peak at 2083 days, but still with a FAP of 0.15\%.  We
are also able to recover the \citet{fischer02} parameters of 47~UMa~c from
the Lick data alone.  Since the total time coverage of the Lick data
presented in \citet{fischer02} is 5114 days (14~yr), an analysis of those
data alone is not adequate to fully constrain the 7586-day period obtained
by our best-fit model which adds four years to the total duration of
observations.  It is possible that the shorter period for 47~UMa~c
reported by \citet{fischer02} is an alias of the true period; at present,
our fits indicate that period to be about three times longer, but with
substantial uncertainty.

To further test the methods by which we conclude that the parameters of
47~UMa~c reported by \citet{fischer02} are dubious, we performed some Monte
Carlo simulations.  From each of the three data sets considered in the fits
described above, we generated 1000 simulated sets of velocities consisting
of a Keplerian signal plus 7 \ms\ of Gaussian noise for each of the two
planets.  The parameters for the inner planet were those listed in Table 2,
and those of the outer planet were those of \citet{fischer02}.  These
simulated datasets retained the times of observation and the error bars
of the originals.  The simulated data were then fit with a one-planet model
exactly as described above, then the residuals of the one-planet fit were
examined by the periodogram method, to determine whether the signal of the
second planet was recovered.  The criteria for recovery were that the
period of the second planet had to be detected correctly and with a FAP of
less than 0.1\%.  This FAP was computed using the analytic FAP formula of
\citet{hb86}.  Of the 1000 trials, only 6 did not result in a successful
recovery of the signal of the second planet.  The correct period was
recovered 999 times, and the FAP exceeded 0.1\% only 5 times; the worst FAP
was 0.24\%.  For comparison, the analytic FAP of the 2212-day peak in the
residuals of the 1-planet fit is 1.7\%, a factor of 7 higher.  These
results indicate that our method should have been able to detect the signal
of 47~UMa~c, had it been present with the parameters given by
\citet{fischer02} and \citet{butler06}.  We conclude that while an
additional long-period object may be present, the data currently available
do not provide sufficient evidence for an orbital solution.

\subsection{14 Her (=HD~145675): Evidence for an outer companion}

Thirty-five radial-velocity measurements of 14~Her (=HD~145675)  obtained
at McDonald Observatory were combined with published data from Keck HIRES
\citep{butler06} and ELODIE \citep{naef04}.  The fit to the combination of
these three data sets is shown in Figure~3, and the system parameters
implied by that fit are given in Table 2.  It is evident from Figure~3
that the single Keplerian fit is inadequate (rms=13.0 \ms); indeed, the
residuals to the fit show a clear curvature. 

\citet{naef04} noted an upward linear velocity trend of 3.6$\pm$0.3 m
s$^{-1}$ yr$^{-1}$ in their observations which covered the time interval
1994 to 2003, but the Keck data of \citet{butler03} indicated no such
trend.  When the complete 12~yr duration of observations is examined, we
see that those Keck data were serendipitously obtained during a period of
relatively constant residual velocity, and the more recent McDonald data
are now indicating a downward trend.  Fitting a double-Keplerian model
(Figure~4) with $T_0$ for the outer body fixed at 2449100.0 yields an
eccentricity consistent with zero: $e=0.02\pm 0.06$.  Fixing the
eccentricity of 14~Her~c at 0.0 gives a minimal orbit solution with period
$P=6906\pm 70$ days, and a velocity semi-amplitude $K=24.5\pm 1.4$ \ms.  
These orbital elements imply a minimum mass M~sin~$i$=2.1 \Mjup\ at a
semimajor axis of 6.9~AU.  This represents the lowest possible mass and
the shortest period for the outer companion.  The rms scatter about the
three data sets is as follows: ELODIE -- 10.4 \ms, Keck -- 2.9 \ms,
McDonald -- 5.8 \ms.  The total rms about this two-planet fit is 8.4 \ms,
with a $\chi^2_{\nu}$ of 1.67.  The minimal orbit proposed above for
14~Her~c has a periastron distance of 6.2~AU, and 14~Her~b has an apastron
distance of 3.76~AU.  The orbit of 14~Her~c, at a large semimajor axis
with a small eccentricity, is similar to that of HD~72659b
\citep{butler03} and HD~50499b \citep{vogt05}.  We emphasize that such
fits are preliminary, and are only given in order to place a lower limit
on the period and mass of this object.  \citet{polishguys} performed
dynamical simulations of the 14~Her system, and proposed the existence of
14~Her~c based on fits to data from \citet{butler03} and \citet{naef04}.  
The two most-favored models had 14~Her~c near the 3:1 or 6:1 mean-motion
resonance (MMR), which coincided with highly stable orbital configurations
as determined by their dynamical simulations.  The preliminary fit
obtained in this work suggests a 4:1 resonance for 14~Her~c.  We attempted
to fit a double-Keplerian model with 14~Her~c fixed at the 3:1 MMR
parameters indicated in \citet{polishguys}, but obtained a substantially
worse fit ($\chi^2_{\nu}=2.2$).

An upper limit on the mass of 14~Her~c can be estimated from the results
of adaptive-optics (AO) imaging by \citet{luhman02}, who used the Keck~II
AO system to search for companions to 25 extrasolar planet-host stars.  
No candidate companions were found around 14~Her, and the detection
limits derived from their study exclude stellar-mass objects objects at
orbital separations $\gtsimeq$ 9~AU (see Luhman \& Jayawardhana 2002,
their Fig.~10).  A similar study by \citet{patience02} using the Lick AO
system also excluded stellar-mass objects beyond about 12.7~AU.  We can
therefore use the upper limit provided by \citet{luhman02} and our lower
bound to constrain the mass of the outer companion between about 2.1 and
80 \Mjup.  A more definitive statement on its nature, of course, requires
many more years of observations, until a second velocity turnaround is
confirmed.  The large separation implied by these data make 14~Her an
attractive target for future direct imaging attempts and for astrometric
follow-up.  The astrometric perturbation due to 14~Her~c would be $\alpha
=$0.9~mas, using the minimal orbit solution given above.  For
sin~$i$=0.5, this signal would be $\alpha =$1.8~mas, equivalent to that
of $\epsilon$~Eri~b \citep{fritz}.  Since the astrometric perturbation
increases with semimajor axis and planet mass, the signal could be much
larger \citep{sozzetti05}.

\subsection{Revised orbital parameters for 7 additional systems}

The addition of new data from the HJS Telescope presented in this paper,
and the use of multiple independent data sets in fitting Keplerian orbits,
have generally improved the precision of the derived planetary parameters
by 25-50\%.  In particular, the precision of the orbital periods have been
improved by the addition of new data, due to the increased number of
orbits now observed.  In this section, we briefly describe the results of
our combined fits for seven additional systems for which the revised
orbital solutions are in agreement with previously published results.  
Our parameters are within 2$\sigma$ of previous values except for $T_0$
and $\omega$ of $\upsilon$~And~c, and $P$ and $\omega$ of
$\upsilon$~And~d, which differ from \citet{butler06} by between 2.7 and
3.8 sigma.

The inner planet of HD~217107, with a period of 7.1 days, was first
reported in \citet{fischer99}, and additional CORALIE data supporting this
discovery were published by \citet{naef01}.  A shallow parabolic trend in
the residuals was noted by \citet{fischer01}.  \citet{vogt05} used more
recent Keck data to postulate an outer companion with $P>7$ years.  
However, this object has not completed a full orbit, and hence there is a
wide range of possible solutions.  \citet{wright05} revised the period of
HD~217107c to 20,140 days (55 years), and \citet{vogt05} give a period of
3150 days; our fits with the short period improved $\chi^2_{\nu}$ by only
0.07 over the long period.  We fit four datasets for HD~217107: Lick
\citep{fischer99}, CORALIE \citep{naef01}, Keck \citep{vogt05}, and 20
observations from McDonald.  We fit a double-Keplerian model using the
parameters for planet~c from \citet{vogt05} as a starting point for the
least-squares fitting procedure.  The rms of this double-Keplerian fit is
9.0 \ms, and the results are given in Table 2.  The resulting parameters
are in agreement with \citet{vogt05} to within 2$\sigma$.  \citet{vogt05}
performed separate fits to the Lick and Keck data and commented on the
high degree of uncertainty in the parameters of HD~217107c.  The combined
fits given in this work support the $\sim$3150-day period reported by
\citet{vogt05}, but the uncertainties remain large.  At present, no
definitive statements can be made until this object completes a
significant fraction of an orbit, or until such time as it can be detected
via direct imaging or astrometry, techniques which are well-suited to very
long-period objects.

The three planets around $\upsilon$ And (=HD~9826) were fit with a
triple-Keplerian model, combining the Lick data of \citet{butler06} and
the Advanced Fiber Optic Echelle (AFOE) data of \citet{butler99} with 41
observations from McDonald.  Lick data preceding the Hamilton spectrometer
upgrade (1995 February) were excluded following \citet{butler97}.  The
total rms about the fit is 14.6 \ms.  As shown in Table 4, analysis of the
residuals to a triple-Keplerian fit resulted in only a marginally
significant periodicity at 2000 days (FAP=0.4\%).

\citet{marcy99} first detected the inner companion to HD~168443 using the
HIRES spectrograph on Keck~I.  The outer 4.8-yr planet was then reported
in \citet{marcy01} and confirmed by \citet{udry02}.  A two-planet fit to
the McDonald data combined with that of \citet{butler06} and
\citet{udry02} yields an rms of 9.5 \ms.  The parameters given in Table~2
are in agreement with those of \citet{butler06} within 2$\sigma$.  
Periodogram analysis of the residuals to the double-Keplerian fit revealed
a 63-day periodicity with a FAP of 0.12\% (Table 4).  Performing the same
analysis on the three datasets separately, however, showed no enhanced
power at that period.  Additionally, the period of the inner planet is 58
days, preventing any object in a 63-day orbit.

The following five planets showed no residual periodicities of interest.
The hot Jupiter orbiting HD~179949 was found by the Anglo-Australian
Planet Search program using the 3.9m Anglo-Australian Telescope (AAT)  
\citep{tinney01}.  The rms about the combined fit is 11.5 \ms, and the
fitted parameters agree with those of \citet{butler06} to within
1$\sigma$.  For 16~Cyg~B (=HD~186427), data from the discovery paper
\citep{cochran97} and Lick data from \citet{butler06} were combined with
37 additional measurements from McDonald Observatory.  The rms about the
combined fit is 10.6 \ms, and the orbital elements are within 2$\sigma$ of
those given in \citet{butler06}.  For HD 195019b, Lick data from
\citet{butler06} were combined with 19 McDonald observations.  The rms
about the combined fit is 15.8 \ms, and the parameters agree with those of
\citet{butler06} within 2$\sigma$.  For HD 210277b, data from
\citet{butler06} and \citet{naef01} were combined with 21 measurements
from McDonald.  The rms about the combined fit is 6.8 \ms, and the
planetary parameters agree with those of \citet{butler06} to within
2$\sigma$.

\section{Summary}

We have combined new radial-velocity observations from McDonald
Observatory with previously published data to improve the precision of the
orbital parameters for 9 planetary systems.  The largest data set yet
assembled on 47~UMa indicates no statistically significant signal
attributable to a second planet in the system.  We have examined the
residuals to our Keplerian fits, and for the case of 14~Her, we find clear
evidence for a distant outer companion, which appears to be in a 4:1
resonance with the inner planet.

\acknowledgements

This research was supported by NASA grants NNG04G141G and NNG05G107G.  
R.W.~acknowledges support from the Sigma Xi Grant-in-Aid of Research and
the Texas Space Grant Consortium.  We are grateful to Barbara McArthur for
her assistance with GaussFit software, and to G.~F.~Benedict for helpful
discussions on astrometry.  This research has made use of NASA's
Astrophysics Data System (ADS), and the SIMBAD database, operated at CDS,
Strasbourg, France.  The Hobby-Eberly Telescope (HET) is a joint project
of the University of Texas at Austin, the Pennsylvania State University,
Stanford University, Ludwig-Maximilians-Universit{\" a}t M{\" u}nchen, and
Georg-August-Universit{\" a}t G{\" o}ttingen The HET is named in honor of
its principal benefactors, William P.~Hobby and Robert E.~Eberly.


\clearpage

\begin{deluxetable}{llllll}
\tabletypesize{\scriptsize}
\tablecolumns{6}
\tablewidth{0pt}
\tablecaption{Stellar Parameters \label{tbl-1}}
\tablehead{
\colhead{Star} & \colhead{Spec.~Type} & \colhead{Mass (\Msun)} & 
\colhead{[Fe/H]} & \colhead{$T_{eff}$ (K)} & \colhead{log$R'_{HK}$}
}

\startdata
$\upsilon$ And & F8V & 1.30 & 0.13$\pm$0.08 & 6212$\pm$64 & -5.01 \\
47 UMa & G1V & 1.07 & 0.06$\pm$0.03 & 5954$\pm$25 & -5.04 \\
14 Her & K0V & 0.90 & 0.43$\pm$0.08 & 5311$\pm$87 & -5.06 \\
HD 168443 & G5 & 0.96 & 0.06$\pm$0.05 & 5617$\pm$35 & -5.14 \\
HD 179949 & F8V & 1.28 & 0.22$\pm$0.05 & 6260$\pm$43 & -4.75 \\
16 Cyg B & G3V & 0.99 & 0.08$\pm$0.04 & 5772$\pm$25 & -5.03 \\
HD 195019 & G3IV-V & 1.06 & 0.08$\pm$0.04 & 5842$\pm$13 & -4.89 \\
HD 210277 & G0V & 0.92 & 0.19$\pm$0.04 & 5532$\pm$14 & -5.11 \\
HD 217107 & G8IV & 1.02 & 0.37$\pm$0.05 & 5646$\pm$26 & -5.13 \\
\enddata
\end{deluxetable}

\begin{deluxetable}{lr@{$\pm$}lr@{$\pm$}lr@{$\pm$}lr@{$\pm$}lr@{$\pm$}lr@{$\pm$}lr@{$\pm$}l} 
\tabletypesize{\scriptsize}
\tablecolumns{8}
\tablewidth{0pt}
\tablecaption{Keplerian Orbital Solutions \label{tbl-2}}
\tablehead{
\colhead{Planet} & \multicolumn{2}{c}{Period} & \multicolumn{2}{c}{$T_0$} 
&
\multicolumn{2}{c}{$e$} & \multicolumn{2}{c}{$\omega$} & 
\multicolumn{2}{c}{K } & \multicolumn{2}{c}{M sin $i$ } & 
\multicolumn{2}{c}{$a$ }\\
\colhead{} & \multicolumn{2}{c}{(days)} & \multicolumn{2}{c}{(JD-2400000)} 
& 
\multicolumn{2}{c}{} & 
\multicolumn{2}{c}{(degrees)} & \multicolumn{2}{c}{(\ms)} & 
\multicolumn{2}{c}{(\Mjup)} & \multicolumn{2}{c}{(AU)} 
 }

\startdata
$\upsilon$ And b & 4.61708 & 0.00006 & 50001.8 & 0.4 & 
0.029 & 0.013 & 46 & 29 & 71.1 & 1.0 & 0.69 & 0.03 & 
0.059 & 0.001 \\
$\upsilon$ And c & 241.52 & 0.21 & 50149.7 & 3.3 & 0.254 & 0.016 & 
232.4 & 4.9 & 56.1 & 1.2 & 1.98 & 0.09 & 0.83 & 0.01 \\
$\upsilon$ And d & 1274.6 & 5.0 & 50074 & 16 & 0.242 & 0.017 & 
258.5 & 5.4 & 64.1 & 1.1 & 3.95 & 0.16 & 2.51 & 0.04 \\
47 UMa b & 1083.2 & 1.8 & 50173 & 65 & 0.049 & 0.014 & 
111 & 22 & 49.3 & 1.0 & 2.60 & 0.13 & 2.11 & 0.04 \\
47 UMa c\tablenotemark{a} & 7586 & 727 & 52134 & 146 & 
\multicolumn{2}{c}{0.005 (fixed)} &
\multicolumn{2}{c}{127 (fixed)} & 13.3 & 1.4 & 1.34 & 0.22 & 7.73 & 0.58 
\\
14 Her b & 1773.4 & 2.5 & 51372.7 & 3.6 & 0.369 & 0.005 & 
22.6 & 0.9 & 90.0 & 0.5 & 4.64 & 0.19 & 2.77 & 0.05 \\
HD 168443 b & 58.1112 & 0.0009 & 50047.45 & 0.04 & 0.530 & 0.001 & 
172.7 & 0.2 & 476.0 & 1.0 & 7.48 & 0.27 & 0.290 & 0.005 \\
HD 168443 c & 1765.8 & 2.2 & 50255 & 4 & 0.222 & 0.003 & 
64.6 & 0.8 & 299.2 & 1.0 & 16.87 & 0.64 & 2.84 & 0.05 \\
HD 179949 b & 3.09250 & 0.00003 & 51793.98 & 0.29 & 0.022 & 0.014 
& 183 & 34 & 112.8 & 1.6 & 0.95 & 0.04 & 0.045 & 0.001 \\
16 Cyg B b & 799.5 & 0.6 & 50539.3 & 1.6 & 0.689 & 0.011 & 
83.4 & 2.1 & 51.2 & 1.1 & 1.68 & 0.07 & 1.68 & 0.03 \\
HD 195019 b & 18.2008 & 0.0003 & 50033.1 & 0.8 & 0.014 & 0.004 
& 239 & 16 & 272.8 & 1.0 & 3.67 & 0.13 & 0.138 & 0.002 \\
HD 210277 b & 442.1 & 0.4 & 50988.2 & 1.6 & 0.472 & 0.011 & 
118.2 & 1.9 & 39.5 & 0.5 & 1.23 & 0.05 & 1.10 & 0.02 \\
HD 217107 b & 7.12689 & 0.00005 & 49998.50 & 0.04 & 0.132 & 0.005 & 
22.7 & 2.0 & 140.6 & 0.7 & 1.33 & 0.05 & 0.073 & 0.001 \\
HD 217107 c & 3352 & 157 & 50921 & 84 & 0.537 & 0.026 &
\multicolumn{2}{c}{164 (fixed)} & 39.8 & 6.4 & 2.50 & 0.48 & 4.41 & 0.21 
\\
\enddata
\tablenotetext{a}{Results for 47 UMa from Model 3 in Table 3.}
\end{deluxetable}

\begin{deluxetable}{lrrrr}
\tabletypesize{\scriptsize}
\tablecolumns{5}
\tablewidth{0pt}
\tablecaption{47 UMa Orbital Solutions \label{tbl-3}}
\tablehead{
\colhead{Parameter} & \colhead{Model 1} & \colhead{Model 
2\tablenotemark{a}} & 
\colhead{Model 3\tablenotemark{b}} & \colhead{Model 4\tablenotemark{c}}}
\startdata
$P_b$ (days) & 1078.3 & 1089.0 & 1083.2 & 1073.1 \\
$T_b$ (JD-2400000) & 52391 & 50356 & 50173 & 50148 \\
$e_b$ & 0.088 & 0.061 & 0.049 & 0.028 \\
$\omega_b$ (degrees) & 127 & 171.8 & 111 & 100 \\ 
$K_b$ (\ms) & 46.7 & 49.3 & 49.3 & 50.2 \\
\hline
$P_c$ (days) & \nodata & 2594 & 7586 & 2594 \\
$T_c$ (JD-2400000) & \nodata & 51363.5 & 52134 & 51363.5 \\
$e_c$ & \nodata & 0.005 & 0.005 & 0.005 \\
$\omega_c$ (degrees) & \nodata & 127 & 127 & 127 \\
$K_c$ (\ms) & \nodata & 11.1 & 13.3 & 11.1 \\
\hline
rms (\ms) & 10.4 & 12.7 & 8.4 & 10.2 \\
$\chi^{2}_{\nu}$ & 3.23 & 5.81 & 2.11 & 2.68 \\
\enddata
\tablenotetext{a}{All parameters fixed at those of \citet{fischer02}}
\tablenotetext{b}{All parameters free except for $e_c$ and $\omega_c$}
\tablenotetext{c}{Only parameters for planet c fixed at those of 
\citet{fischer02}}
\end{deluxetable}

\begin{deluxetable}{lrr}
\tabletypesize{\scriptsize}
\tablecolumns{3}
\tablewidth{0pt}
\tablecaption{Periodogram Analysis \label{tbl-4}}
\tablehead{
\colhead{Star} & \colhead{Period (days)} & \colhead{FAP\tablenotemark{a}}}
\startdata
$\upsilon$ And & 2000.0 & 0.004 \\
47 UMa & 2212 & 0.272 \\
14 Her\tablenotemark{b} & 2.37 & 0.118 \\
HD 168443 & 63.37 & 0.001 \\
HD 179949 & 32.05 & 0.023 \\
16 Cyg B & 2.32 & 0.895 \\
HD 195019 & 41.60 & 0.045 \\
HD 210277 & 15.50 & 0.392 \\
HD 217107\tablenotemark{b} & 485.44 & 0.332 \\
\enddata
\tablenotetext{a}{10000 bootstraps.}
\tablenotetext{b}{Residuals obtained from 2-planet fit.}
\end{deluxetable}

\begin{deluxetable}{lrr}
\tabletypesize{\scriptsize}
\tablecolumns{3}
\tablewidth{0pt}
\tablecaption{HJS Telescope Radial Velocities for $\upsilon$ And (=HD~9826) 
\label{tbl-5}}
\tablehead{ 
\colhead{JD-2400000} & \colhead{Velocity (\ms)} & \colhead{Uncertainty
(\ms)}}
\startdata
51452.86152  &    -13.8  &    9.2  \\
51452.86540  &     -3.0  &   13.4  \\
51504.69248  &      0.8  &    8.1  \\
51530.78235  &    -22.2  &    9.4  \\
51532.65475  &    -16.0  &   10.8  \\
51557.64589  &     26.8  &    7.7  \\
51750.96883  &    113.1  &   10.6  \\
51775.91401  &    -54.7  &    8.1  \\
51778.93251  &     63.8  &    9.8  \\
51809.77554  &    -62.0  &    8.6  \\
51859.76793  &    -50.3  &    6.8  \\
51861.85297  &     96.7  &    7.1  \\
51861.85893  &     96.1  &    7.7  \\
51862.79111  &     91.5  &   11.9  \\
51862.79735  &     68.8  &   10.1  \\
51918.75978  &      5.0  &    8.7  \\
51918.76434  &     23.1  &    9.5  \\
51920.67309  &     46.5  &    6.9  \\
51946.68903  &     23.8  &   14.3  \\
51987.57597  &      6.3  &    9.1  \\
52142.87690  &     55.5  &    9.3  \\
52218.75362  &    -63.5  &    9.6  \\
52249.68686  &     -6.3  &    8.5  \\
52329.57235  &   -125.9  &    8.7  \\
52495.93509  &   -151.8  &   14.9  \\
52539.84157  &   -104.7  &    9.3  \\
52539.84425  &   -113.3  &    9.4  \\
52577.85533  &      5.0  &    8.5  \\
52619.75862  &     44.4  &    8.5  \\
52933.79237  &    102.3  &   10.4  \\
52933.79503  &     92.7  &    9.8  \\
52958.68304  &     -5.6  &    9.1  \\
53015.67064  &     28.9  &    8.2  \\
53035.57199  &    -20.0  &    7.0  \\
53394.63389  &     77.3  &   10.2  \\
53394.64950  &     85.3  &   10.0  \\
53394.65545  &     76.5  &   10.8  \\
53632.91257  &   -106.3  &    9.2  \\
53690.78040  &    -29.4  &   10.0  \\
53691.77259  &   -130.3  &   13.4  \\
53746.71222  &   -150.9  &    8.3  \\
\enddata
\end{deluxetable}

\begin{deluxetable}{lrr}
\tabletypesize{\scriptsize}
\tablecolumns{3}
\tablewidth{0pt}
\tablecaption{HJS Telescope Radial Velocities for 47 UMa (=HD~95128)
\label{tbl-6}}
\tablehead{
\colhead{JD-2400000} & \colhead{Velocity (\ms)} & \colhead{Uncertainty
(\ms)}}
\startdata
51010.62898  &     51.6  &    6.4  \\
51212.97474  &    -10.8  &    5.5  \\
51240.81250  &     -7.3  &    6.2  \\
51274.78993  &    -11.2  &    5.4  \\
51326.70558  &    -22.4  &    6.2  \\
51504.95996  &    -42.2  &    6.0  \\
51530.01978  &    -29.4  &    6.9  \\
51555.94972  &    -25.5  &    5.8  \\
51655.74023  &      5.8  &    5.8  \\
51686.75156  &     -6.7  &    6.5  \\
51750.60418  &      2.0  &    6.6  \\
51861.01895  &     53.7  &    6.9  \\
51917.93086  &     47.5  &    7.1  \\
51987.85527  &     47.8  &    8.5  \\
52004.83235  &     59.7  &    6.0  \\
52039.77936  &     54.8  &    7.5  \\
52116.60554  &     39.4  &    7.6  \\
52249.00010  &      9.5  &    7.6  \\
52303.89238  &     -9.6  &    5.5  \\
52305.84757  &    -11.8  &    6.1  \\
52327.86285  &     12.3  &   16.6  \\
52353.85949  &    -12.9  &    7.7  \\
52661.95399  &    -24.9  &    5.4  \\
53017.93695  &     61.5  &    7.5  \\
53069.76686  &     60.7  &    6.4  \\
53692.03243  &    -49.7  &    8.1  \\
53748.89147  &    -47.9  &    6.0  \\
53787.91198  &    -35.2  &    6.3  \\
53805.88756  &    -29.3  &    5.6  \\
53809.80777  &    -30.8  &    6.1  \\
53805.88756  &    -29.3  &    5.6  \\
53809.80777  &    -30.8  &    6.1  \\
53787.91198  &    -35.2  &    6.3  \\
53861.74397  &    -17.1  &    6.1  \\
53909.61977  &     13.8  &    7.0  \\
\enddata
\end{deluxetable}

\begin{deluxetable}{lrr}
\tabletypesize{\scriptsize}
\tablecolumns{3}
\tablewidth{0pt}
\tablecaption{HET Radial Velocities for 47 UMa (=HD~95128)
\label{tbl-7}}
\tablehead{
\colhead{JD-2400000} & \colhead{Velocity (\ms)} & \colhead{Uncertainty
(\ms)}}
\startdata
53313.99225  &     61.3  &    1.5  \\
53313.99417  &     64.1  &    1.6  \\
53313.99608  &     52.5  &    1.7  \\
53314.99012  &     56.2  &    1.3  \\
53314.99203  &     59.9  &    1.3  \\
53317.98891  &     48.4  &    1.6  \\
53317.99082  &     48.4  &    1.4  \\
53317.99273  &     47.5  &    1.4  \\
53334.94873  &     48.2  &    1.4  \\
53334.95296  &     50.9  &    1.4  \\
53334.95591  &     54.0  &    1.4  \\
53335.94413  &     52.8  &    1.4  \\
53335.94708  &     54.5  &    1.4  \\
53335.95003  &     54.5  &    1.3  \\
53338.92569  &     57.9  &    2.0  \\
53338.92756  &     53.0  &    3.1  \\
53338.92947  &     46.4  &    2.1  \\
53338.93800  &     45.0  &    1.9  \\
53338.93991  &     48.6  &    1.9  \\
53338.94181  &     57.3  &    1.8  \\
53338.94426  &     52.2  &    1.9  \\
53338.94617  &     46.5  &    1.8  \\
53338.94808  &     47.7  &    1.9  \\
53340.91533  &     58.2  &    1.4  \\
53340.91724  &     53.0  &    1.5  \\
53340.91915  &     58.0  &    1.4  \\
53346.92015  &     50.0  &    1.5  \\
53346.92207  &     50.6  &    1.4  \\
53346.92398  &     51.7  &    1.3  \\
53348.90749  &     51.3  &    1.6  \\
53348.90939  &     56.8  &    1.6  \\
53348.91131  &     52.4  &    1.5  \\
53350.91699  &     51.9  &    1.4  \\
53357.87818  &     39.1  &    1.8  \\
53357.88009  &     50.8  &    1.8  \\
53357.88200  &     40.8  &    1.6  \\
53359.87351  &     46.6  &    2.2  \\
53359.87542  &     43.8  &    2.1  \\
53359.87733  &     48.4  &    2.1  \\
53365.86302  &     39.9  &    1.9  \\
53365.86489  &     48.7  &    2.1  \\
53365.86678  &     47.2  &    2.0  \\
53367.86198  &     49.4  &    1.8  \\
53367.86389  &     41.8  &    1.7  \\
53367.86580  &     43.7  &    1.8  \\
53371.85542  &     32.8  &    1.8  \\
53371.85733  &     41.6  &    2.5  \\
53371.85925  &     41.8  &    1.7  \\
53373.85759  &     40.9  &    2.1  \\
53373.85950  &     32.0  &    4.0  \\
53389.79570  &     42.8  &    2.0  \\
53389.79762  &     37.9  &    2.0  \\
53389.79953  &     35.2  &    1.9  \\
53391.79094  &     35.3  &    2.0  \\
53391.79285  &     36.6  &    2.1  \\
53391.79477  &     34.5  &    1.8  \\
53395.77629  &     42.2  &    1.8  \\
53395.77819  &     32.7  &    1.8  \\
53395.78010  &     42.7  &    1.9  \\
53400.99279  &     33.2  &    1.7  \\
53400.99470  &     28.6  &    1.7  \\
53400.99661  &     26.1  &    1.7  \\
53408.76776  &     37.1  &    1.9  \\
53408.76968  &     30.5  &    2.1  \\
53408.77158  &     29.1  &    2.2  \\
53414.72643  &     28.5  &    2.1  \\
53414.72832  &     34.6  &    2.0  \\
53414.73023  &     33.9  &    1.9  \\
53416.70849  &     29.2  &    2.2  \\
53416.71038  &     31.4  &    2.1  \\
53416.71231  &     33.6  &    2.1  \\
53421.93924  &     23.3  &    1.3  \\
53421.94115  &     25.9  &    1.4  \\
53421.94306  &     24.6  &    1.4  \\
53423.70290  &     29.4  &    1.3  \\
53423.70481  &     26.5  &    1.4  \\
53423.70672  &     23.0  &    1.4  \\
53432.90612  &     21.4  &    1.3  \\
53432.90802  &     18.9  &    1.3  \\
53432.90993  &     20.6  &    1.3  \\
53433.90512  &     22.7  &    1.2  \\
53433.90696  &     24.3  &    1.2  \\
53433.90835  &     23.3  &    1.2  \\
53437.65943  &     27.5  &    1.6  \\
53437.66100  &     25.6  &    1.4  \\
53437.66291  &     26.5  &    1.3  \\
53437.66489  &     26.9  &    1.2  \\
53439.65763  &     25.3  &    1.3  \\
53439.65954  &     30.6  &    1.3  \\
53439.66145  &     24.5  &    1.1  \\
53440.89735  &     32.7  &    1.2  \\
53440.90029  &     30.7  &    1.4  \\
53440.90324  &     34.8  &    1.4  \\
53476.80210  &     16.3  &    1.2  \\
53476.80400  &     11.2  &    1.5  \\
53476.80591  &     10.3  &    1.4  \\
53479.77654  &     11.7  &    1.3  \\
53479.77844  &     13.1  &    1.4  \\
53479.78035  &      7.3  &    1.4  \\
53481.76429  &      6.2  &    1.3  \\
53481.76620  &      5.6  &    1.3  \\
53481.76811  &      7.5  &    1.4  \\
53486.77539  &      9.3  &    1.3  \\
53486.77730  &      9.1  &    1.3  \\
53486.77922  &     13.8  &    1.3  \\
53488.76596  &      6.8  &    1.3  \\
53488.76787  &      8.5  &    1.3  \\
53488.76978  &      7.9  &    1.3  \\
53512.68994  &      5.5  &    1.4  \\
53512.69185  &      0.1  &    1.5  \\
53512.69376  &     -0.6  &    1.5  \\
53526.63295  &     -8.0  &    1.5  \\
53526.63492  &     -9.2  &    1.5  \\
53526.63683  &     -8.4  &    1.6  \\
53526.63847  &     -6.1  &    1.6  \\
53526.63969  &    -13.1  &    1.8  \\
53526.64090  &     -5.8  &    1.6  \\
53539.63731  &    -15.7  &    1.6  \\
53539.63922  &    -15.6  &    1.7  \\
53539.64114  &    -22.7  &    1.7  \\
53708.91865  &      8.0  &    1.7  \\
53708.92004  &     12.7  &    1.8  \\
53708.92143  &      4.9  &    1.9  \\
53709.92062  &     10.8  &    1.8  \\
53709.92253  &      0.0  &    1.6  \\
53709.92444  &      6.6  &    1.5  \\
53710.91177  &     13.2  &    1.5  \\
53710.91316  &     13.5  &    1.5  \\
53710.91456  &     11.0  &    1.5  \\
53711.92767  &     17.5  &    2.4  \\
53711.92906  &      9.8  &    2.8  \\
53711.93044  &     10.6  &    2.4  \\
53711.93510  &      8.6  &    1.5  \\
53711.93649  &      7.1  &    1.4  \\
53711.93788  &      4.5  &    1.4  \\
53721.87889  &      3.5  &    1.9  \\
53721.88028  &     18.0  &    1.6  \\
53721.88168  &      6.8  &    1.7  \\
53723.86894  &     11.5  &    1.7  \\
53723.87032  &     12.1  &    1.7  \\
53723.87171  &     11.1  &    1.6  \\
53725.86007  &      9.7  &    1.7  \\
53725.86146  &     19.2  &    1.7  \\
53725.86285  &     20.1  &    1.9  \\
53736.83987  &     -0.9  &    1.9  \\
53736.84137  &      4.3  &    1.8  \\
53736.84288  &      6.0  &    1.8  \\
53738.82611  &     14.7  &    1.6  \\
53738.82750  &      6.7  &    1.7  \\
53738.82890  &     11.5  &    1.7  \\
53734.87673  &      6.1  &    1.8  \\
53734.87812  &      1.5  &    1.9  \\
53734.87951  &      7.9  &    1.9  \\
53742.81869  &     13.3  &    1.8  \\
53742.82008  &      7.7  &    1.9  \\
53742.82147  &     18.8  &    1.8  \\
53743.81885  &      6.8  &    1.9  \\
53743.82024  &     15.2  &    2.0  \\
53743.82163  &      9.8  &    2.0  \\
53744.82153  &      9.0  &    1.5  \\
53744.82292  &     12.2  &    1.6  \\
53744.82431  &      7.6  &    1.6  \\
53751.79848  &      4.3  &    2.1  \\
53751.79987  &     12.2  &    1.9  \\
53751.80126  &      9.3  &    1.9  \\
53746.80595  &      7.8  &    1.8  \\
53746.80758  &      9.0  &    1.8  \\
53746.80920  &     10.9  &    1.8  \\
53757.03611  &     22.9  &    1.7  \\
53757.03749  &     18.2  &    1.8  \\
53757.03887  &     12.6  &    1.8  \\
53771.75959  &     14.6  &    1.8  \\
53771.76109  &      5.0  &    2.1  \\
53771.76259  &      9.5  &    2.1  \\
53775.73900  &     30.9  &    2.0  \\
53775.74040  &     20.8  &    1.9  \\
53775.74179  &     22.6  &    2.0  \\
53777.96474  &     34.3  &    2.0  \\
53777.96664  &     21.8  &    2.2  \\
53777.96855  &     27.1  &    2.1  \\
53779.96267  &     21.7  &    2.0  \\
53779.96405  &     24.2  &    1.9  \\
53779.96543  &     24.8  &    1.9  \\
53786.70391  &      9.7  &    2.1  \\
53786.70737  &     15.6  &    2.1  \\
53786.71085  &     17.4  &    2.2  \\
53795.91621  &     15.5  &    1.6  \\
53795.91760  &     18.7  &    1.6  \\
53795.91899  &     22.8  &    1.6  \\
53795.92042  &     21.4  &    1.7  \\
53795.92181  &     19.4  &    1.7  \\
53795.92320  &      9.6  &    1.9  \\
53797.66582  &     23.9  &    1.7  \\
53797.66773  &     15.6  &    2.2  \\
53797.66964  &     17.5  &    2.2  \\
53894.65375  &     46.4  &    1.3  \\
53894.65514  &     44.4  &    1.4  \\
53894.65653  &     50.4  &    1.5  \\
53901.63954  &     46.8  &    1.5  \\
53901.64093  &     43.2  &    1.5  \\
\enddata
\end{deluxetable}

\begin{deluxetable}{lrr}
\tabletypesize{\scriptsize}
\tablecolumns{3}
\tablewidth{0pt} 
\tablecaption{HJS Telescope Radial Velocities for 14 Her (=HD~145675)
\label{tbl-8}}
\tablehead{
\colhead{JD-2400000} & \colhead{Velocity (\ms)} & \colhead{Uncertainty
(\ms)}}
\startdata
51329.82953  &    154.3  &    7.4  \\
51358.76587  &    148.8  &    6.5  \\
51417.72632  &    127.8  &    7.5  \\
51449.62644  &    106.7  &    6.9  \\
51656.89886  &     11.7  &    7.3  \\
51689.69367  &     -0.5  &    7.2  \\
51750.74058  &     -0.8  &    8.0  \\
51777.71797  &    -20.2  &    6.7  \\
51809.60771  &    -21.1  &    6.8  \\
52037.88607  &    -31.4  &    7.0  \\
52115.73144  &    -35.1  &    7.1  \\
52145.70159  &    -23.4  &    7.7  \\
52181.62915  &    -23.7  &    7.1  \\
52354.98989  &    -25.0  &   12.7  \\
52451.80971  &     -2.1  &    7.4  \\
52454.80637  &     -3.2  &    6.7  \\
52471.72725  &    -12.3  &    6.4  \\
52495.71816  &    -13.9  &    7.4  \\
52541.58879  &     -0.7  &    7.0  \\
52806.74254  &     53.6  &    6.9  \\
52841.82135  &     55.6  &    7.5  \\
52933.55588  &     90.8  &    7.2  \\
53215.73683  &     90.9  &    8.0  \\
53505.84769  &    -31.7  &   10.0  \\
53505.87913  &    -42.9  &    8.2  \\
53563.74131  &    -39.8  &    6.6  \\
53586.66183  &    -31.3  &    8.0  \\
53633.61295  &    -47.0  &    7.2  \\
53636.60040  &    -56.8  &    6.6  \\
53805.95471  &    -71.0  &    7.2  \\
53809.96897  &    -66.3  &    7.2  \\
53840.90354  &    -62.0  &    8.2  \\
53863.84247  &    -62.0  &    9.3  \\
53909.73446  &    -57.6  &    7.3  \\
53927.75413  &    -58.5  &    6.8  \\
\enddata
\end{deluxetable}

\begin{deluxetable}{lrr}
\tabletypesize{\scriptsize}
\tablecolumns{3}
\tablewidth{0pt}
\tablecaption{HJS Telescope Radial Velocities for HD 168443 
\label{tbl-9}}
\tablehead{
\colhead{JD-2400000} & \colhead{Velocity (\ms)} & \colhead{Uncertainty
(\ms)}}
\startdata
51329.88720  &   -523.3  &    6.2  \\
51360.83723  &    161.3  &    5.9  \\
51417.74908  &    208.1  &    6.2  \\
51451.63634  &    -54.8  &    6.2  \\
51689.83601  &    267.1  &    5.9  \\
51752.70677  &    369.2  &    5.8  \\
51776.68347  &    364.6  &    6.7  \\
51810.65202  &    369.8  &    5.8  \\
51861.53311  &    252.0  &    7.5  \\
52040.91217  &    136.4  &    5.7  \\
52116.78446  &     97.1  &    6.5  \\
52453.83556  &   -197.4  &    6.1  \\
52473.72399  &   -233.8  &    6.4  \\
52492.69696  &   -766.5  &    7.6  \\
52540.69172  &   -552.2  &    6.3  \\
52577.57953  &   -156.6  &    5.8  \\
52840.83905  &   -684.6  &    7.1  \\
52932.59046  &      2.3  &    5.7  \\
53504.89012  &    392.3  &    6.5  \\
53565.82109  &    442.8  &    6.8  \\
53863.92552  &     78.8  &    6.8  \\
53911.89301  &     27.4  &    6.6  \\
\enddata
\end{deluxetable}

\begin{deluxetable}{lrr}
\tabletypesize{\scriptsize}
\tablecolumns{3}
\tablewidth{0pt}
\tablecaption{HJS Telescope Radial Velocities for HD~179949
\label{tbl-10}}
\tablehead{
\colhead{JD-2400000} & \colhead{Velocity (\ms)} & \colhead{Uncertainty
(\ms)}}
\startdata
51067.67648  &    -72.4  &    7.7  \\
51121.53727  &    109.3  &   10.2  \\
51328.90372  &     94.1  &    8.3  \\
51360.81419  &   -106.4  &    9.0  \\
51451.60447  &     55.6  &    8.2  \\
51689.88414  &     72.1  &    9.0  \\
51750.78249  &   -121.0  &    9.2  \\
51775.71719  &    -81.5  &    8.6  \\
51812.66687  &    -89.6  &   11.2  \\
52040.93486  &    -62.9  &    8.6  \\
52116.80005  &    104.0  &    9.2  \\
52492.71373  &    -77.8  &    8.7  \\
52577.54345  &     98.7  &    8.4  \\
52840.82950  &    100.4  &   12.2  \\
52933.58832  &    122.3  &    8.0  \\
53565.85370  &   -104.1  &   10.6  \\
53911.90672  &    -41.0  &   10.4  \\
\enddata
\end{deluxetable}

\begin{deluxetable}{lrr}
\tabletypesize{\scriptsize}
\tablecolumns{3}
\tablewidth{0pt}
\tablecaption{HJS Telescope Radial Velocities for 16 Cyg B (=HD~186427) 
\label{tbl-11}}
\tablehead{
\colhead{JD-2400000} & \colhead{Velocity (\ms)} & \colhead{Uncertainty
(\ms)}}
\startdata
51008.86449  &     -7.0  &    6.7  \\
51011.86885  &     -1.9  &    7.0  \\
51065.73704  &      0.5  &    7.7  \\
51121.69528  &     15.6  &    7.9  \\
51154.57100  &     14.7  &    7.3  \\
51328.94395  &     29.1  &    6.7  \\
51361.87182  &    -43.2  &    7.5  \\
51414.67124  &    -48.4  &    7.3  \\
51449.78342  &    -41.3  &    6.8  \\
51502.60158  &    -42.5  &    7.8  \\
51529.57703  &    -36.6  &    8.0  \\
51689.90882  &    -19.1  &    6.9  \\
51753.74228  &      7.5  &    7.6  \\
51777.79311  &     -3.6  &    6.9  \\
51861.66676  &     24.4  &    7.1  \\
52039.94285  &     35.7  &    6.4  \\
52115.84910  &     48.7  &    6.9  \\
52144.77541  &    -16.5  &    7.0  \\
52181.72467  &    -51.1  &    7.3  \\
52451.88739  &    -16.3  &    6.2  \\
52471.85446  &    -15.1  &    6.4  \\
52495.78667  &      2.4  &    9.2  \\
52538.66450  &     -7.3  &    6.8  \\
52577.67137  &     -9.1  &    7.4  \\
52600.58553  &      1.5  &    7.4  \\
52620.56341  &      0.2  &    7.4  \\
52807.88580  &     23.2  &    6.8  \\
52840.96694  &     34.9  &    7.7  \\
52930.68220  &     36.5  &    7.5  \\
52932.66273  &     21.3  &    6.7  \\
52960.59294  &    -27.4  &    7.3  \\
53163.86157  &     -7.0  &    7.0  \\
53321.64999  &     -9.6  &   14.7  \\
53585.84777  &     27.6  &    7.0  \\
53654.68515  &     45.4  &    7.0  \\
53689.59980  &     57.8  &    7.4  \\
53907.87292  &    -24.1  &    7.7  \\
\enddata
\end{deluxetable}

\begin{deluxetable}{lrr}
\tabletypesize{\scriptsize}
\tablecolumns{3}
\tablewidth{0pt}
\tablecaption{HJS Telescope Radial Velocities for HD 195019 
\label{tbl-12}}
\tablehead{
\colhead{JD-2400000} & \colhead{Velocity (\ms)} & \colhead{Uncertainty
(\ms)}}
\startdata
51451.76016  &   -165.6  &    5.6  \\
51503.63673  &   -199.1  &    6.2  \\
51776.76858  &   -221.6  &    4.8  \\
51778.74737  &   -191.2  &    4.5  \\
51860.61320  &    298.7  &    5.3  \\
51862.64533  &    152.9  &    6.3  \\
52114.90052  &    317.7  &    5.9  \\
52221.65789  &    290.2  &    7.1  \\
52472.85791  &    -14.4  &    6.9  \\
52492.82837  &    136.8  &    7.5  \\
52538.72733  &    -74.6  &    6.4  \\
52807.86835  &    268.4  &    6.9  \\
53321.56421  &    -95.5  &   12.3  \\
53505.93331  &   -185.1  &   22.4  \\
53563.87408  &   -103.1  &    5.6  \\
53584.81430  &    159.5  &    7.8  \\
53633.65906  &   -217.5  &    7.7  \\
53636.72312  &    -86.5  &    6.7  \\
53691.64472  &    -69.9  &    6.8  \\
\enddata
\end{deluxetable}

\begin{deluxetable}{lrr}
\tabletypesize{\scriptsize}
\tablecolumns{3}
\tablewidth{0pt}
\tablecaption{HJS Telescope Radial Velocities for HD 210277
\label{tbl-13}}
\tablehead{
\colhead{JD-2400000} & \colhead{Velocity (\ms)} & \colhead{Uncertainty
(\ms)}}
\startdata
51531.59673  &    -30.4  &    5.4  \\
51557.53626  &     13.1  &    6.1  \\
51558.54314  &     -1.3  &    6.3  \\
51689.93883  &     12.3  &    6.3  \\
51750.88078  &     24.3  &    5.9  \\
51776.81672  &     27.7  &    5.3  \\
51860.63668  &      1.9  &   11.7  \\
51917.54532  &    -58.3  &    7.4  \\
52116.88917  &     10.3  &    7.7  \\
52221.67895  &     30.4  &    6.2  \\
52473.83462  &     -9.8  &    5.6  \\
52492.85694  &     -2.7  &    8.2  \\
52540.77428  &      6.2  &    6.4  \\
52621.61830  &      9.1  &    5.3  \\
52840.95838  &    -28.5  &    6.0  \\
52931.71553  &      3.2  &    6.1  \\
53563.94438  &     35.3  &    5.5  \\
53633.72641  &    -26.7  &    5.7  \\
53635.81499  &    -11.6  &    5.9  \\
53689.64535  &    -46.1  &    5.8  \\
53927.90889  &     41.6  &    8.1  \\
\enddata
\end{deluxetable}

\begin{deluxetable}{lrr}
\tabletypesize{\scriptsize}
\tablecolumns{3}
\tablewidth{0pt}
\tablecaption{HJS Telescope Radial Velocities for HD 217107
\label{tbl-14}}
\tablehead{
\colhead{JD-2400000} & \colhead{Velocity (\ms)} & \colhead{Uncertainty
(\ms)}}
\startdata
51449.84000  &    -44.9  &    4.8  \\
51556.57950  &    -52.7  &    5.1  \\
51750.89391  &    154.4  &    5.2  \\
51776.82866  &    -79.6  &    4.9  \\
51777.83705  &      0.8  &    5.7  \\
51778.80954  &     97.5  &    4.6  \\
51778.82275  &    101.5  &    5.3  \\
51809.80381  &     38.2  &    5.0  \\
51860.65078  &    -28.0  &   16.0  \\
51862.66594  &    -75.8  &    6.8  \\
51918.58936  &   -102.3  &    4.7  \\
52116.89988  &    -15.7  &    7.3  \\
52219.68045  &      1.5  &    6.5  \\
52473.84637  &    -83.3  &    5.0  \\
52492.88220  &    157.8  &    8.7  \\
52540.78985  &     46.9  &    5.3  \\
52895.85865  &    -72.5  &    5.0  \\
52931.80527  &    -49.1  &    4.9  \\
53017.56950  &    -29.9  &    6.0  \\
53563.89812  &    -51.8  &    4.8  \\
53630.81136  &     -9.2  &    6.8  \\
53635.82600  &    -81.0  &    6.5  \\
53689.65961  &    177.3  &    5.9  \\
\enddata
\end{deluxetable}

\epsscale{1.00}
\clearpage

\begin{figure}
\plottwo{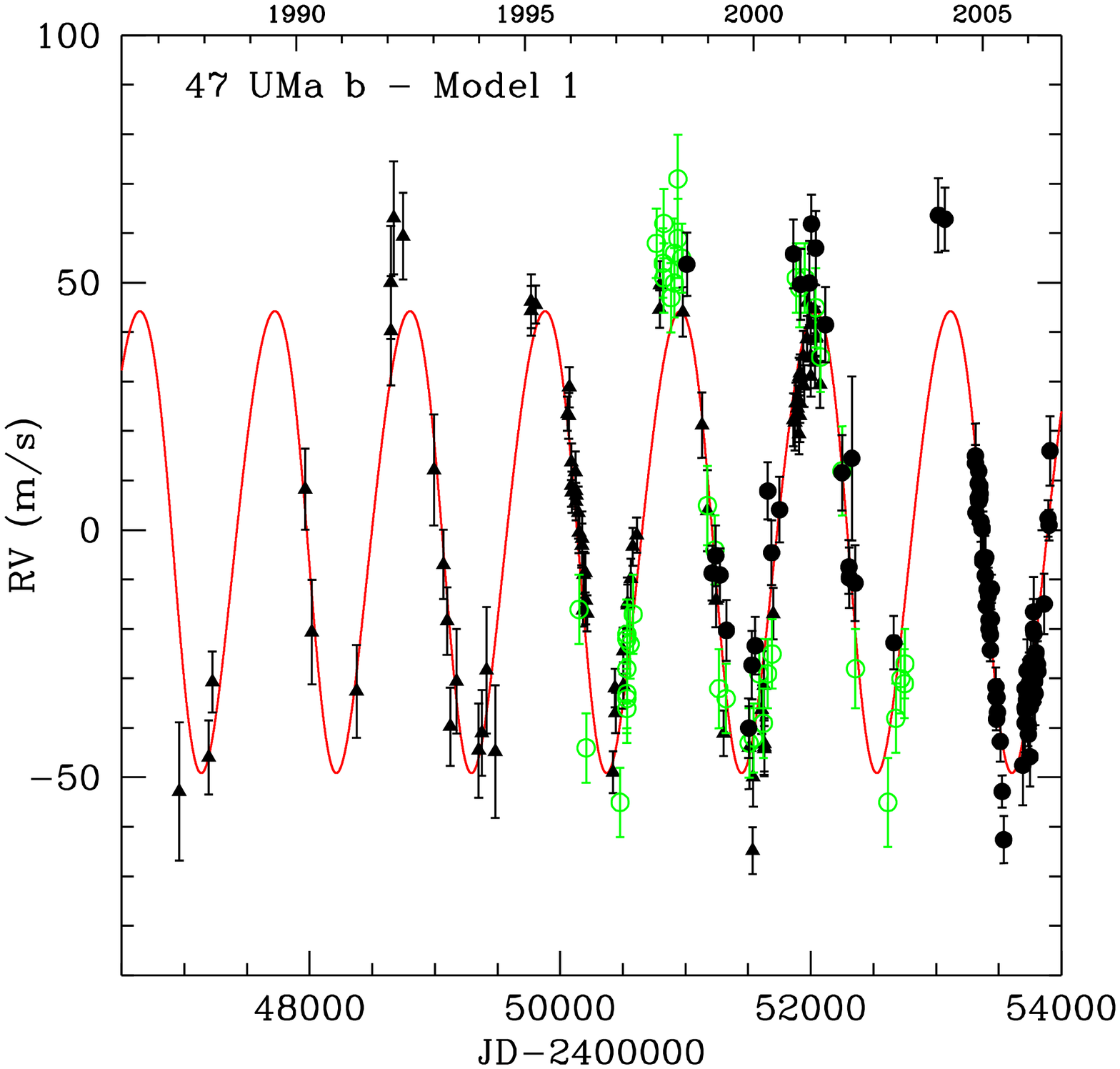}{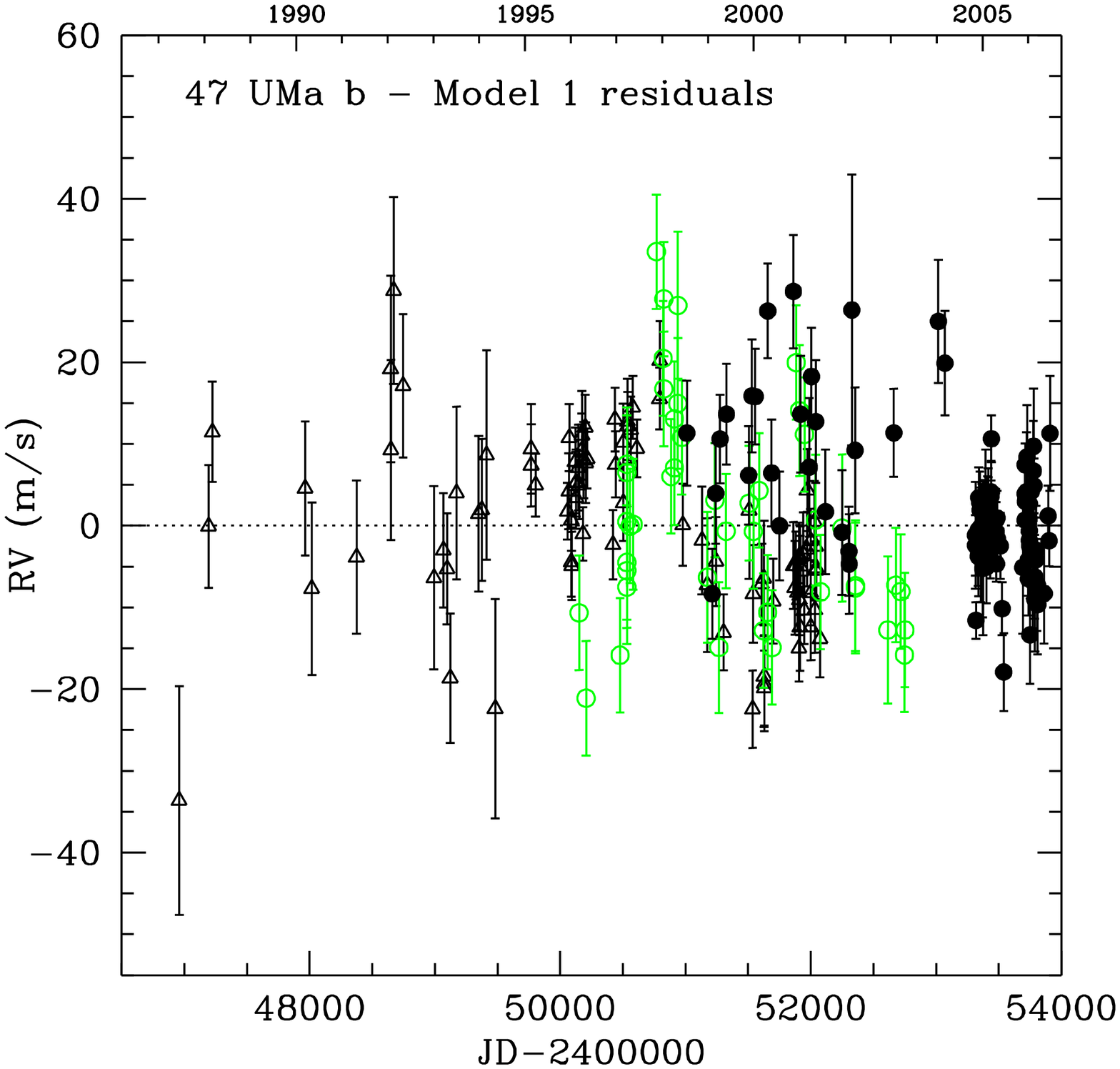}
\caption{Left panel: One-planet orbital fit for 47~UMa~b (Model 1). Open 
triangles
are from Lick \citep{fischer02}, open circles are from ELODIE 
\citep{naef04}, and filled circles are from McDonald (2.7m and HET). 
Right panel: Residuals to the 1-planet fit. }
\end{figure}

\begin{figure}
\plottwo{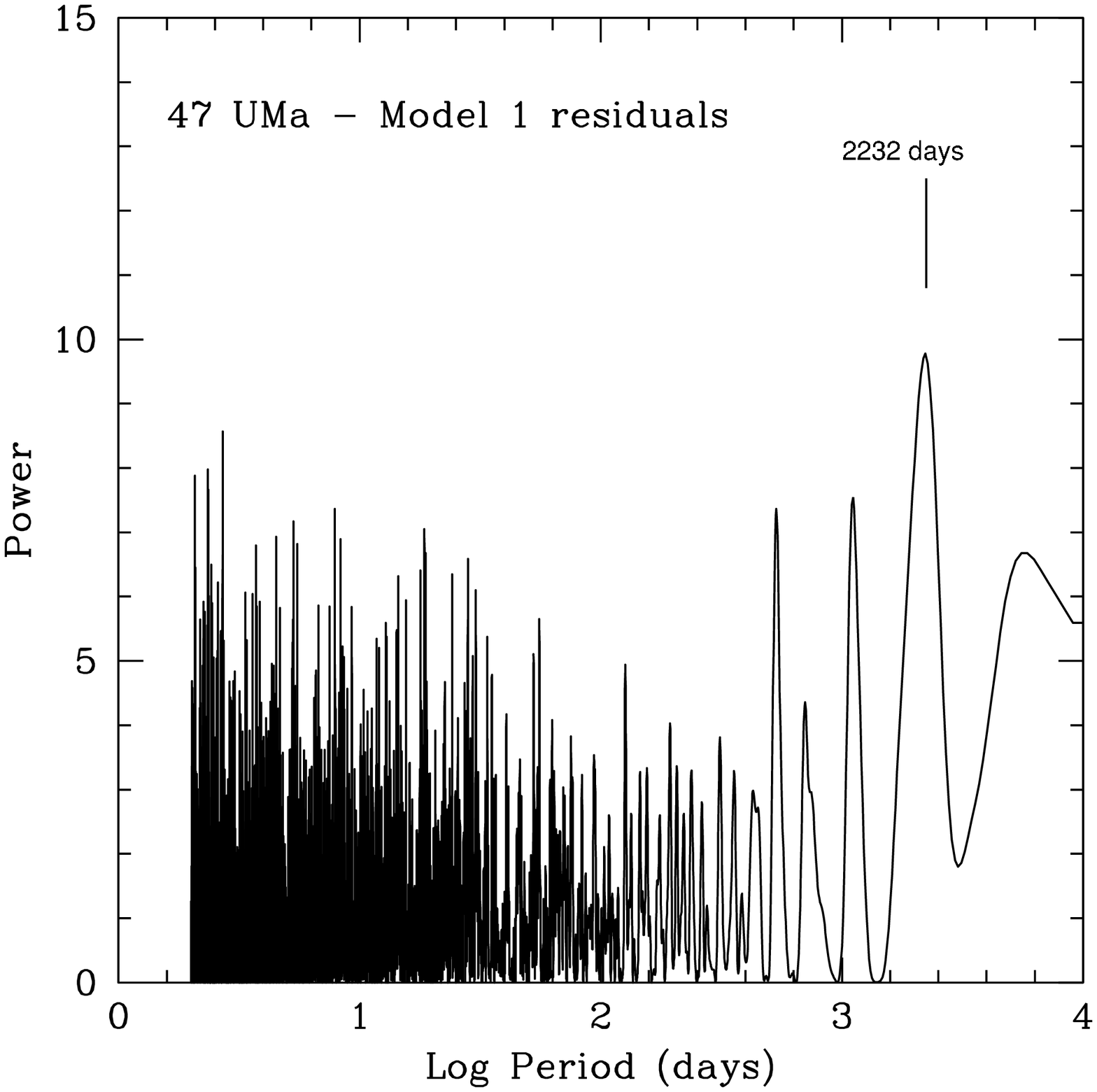}{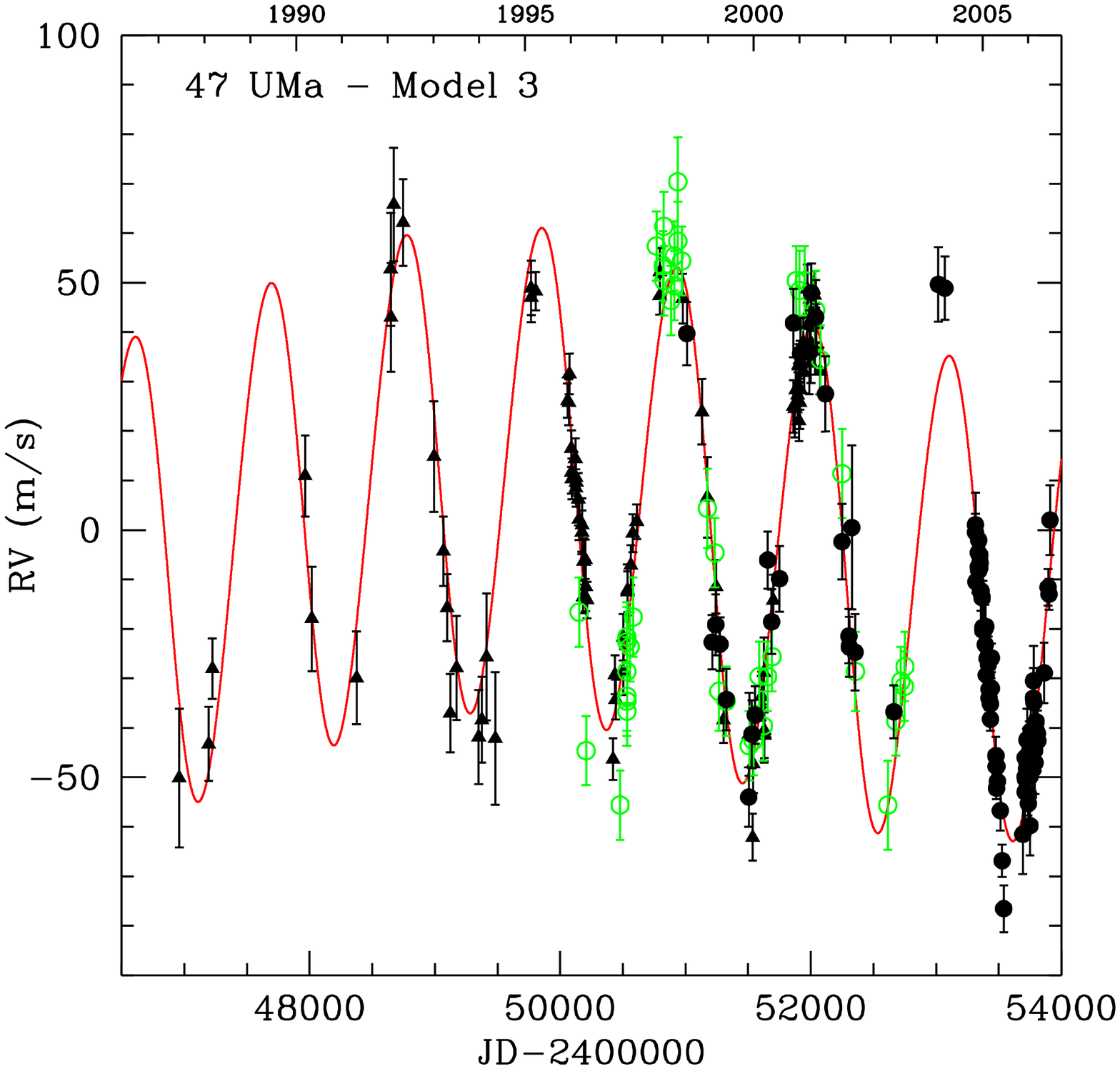}
\caption{Left panel: Lomb-Scargle periodogram for 47~UMa, after removal of 
47~UMa~b.  Right panel: Best-fit double Keplerian 
model for 47~UMa. The symbols have the same meaning as in Fig.~1. } 
\end{figure}

\begin{figure}
\plottwo{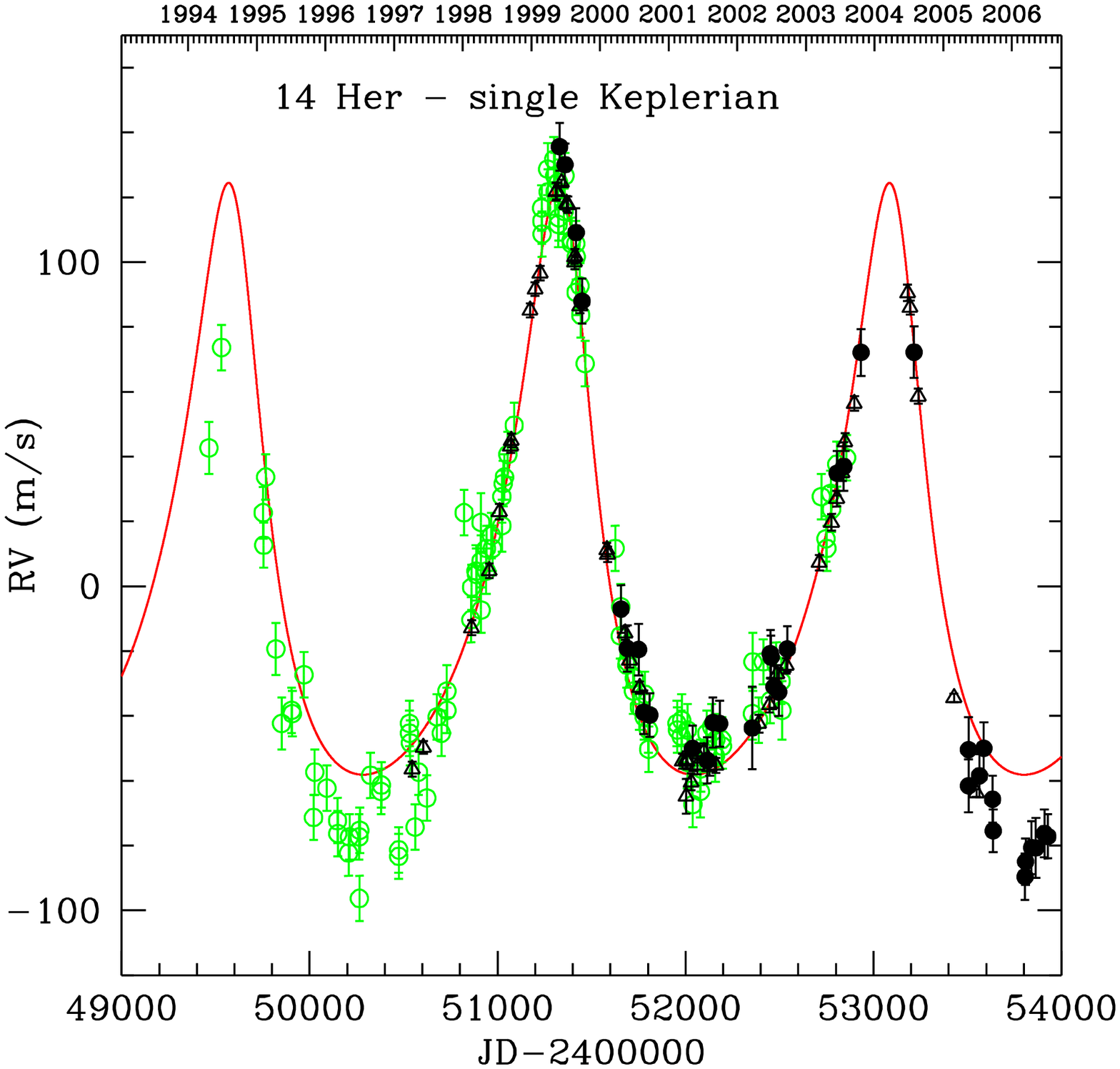}{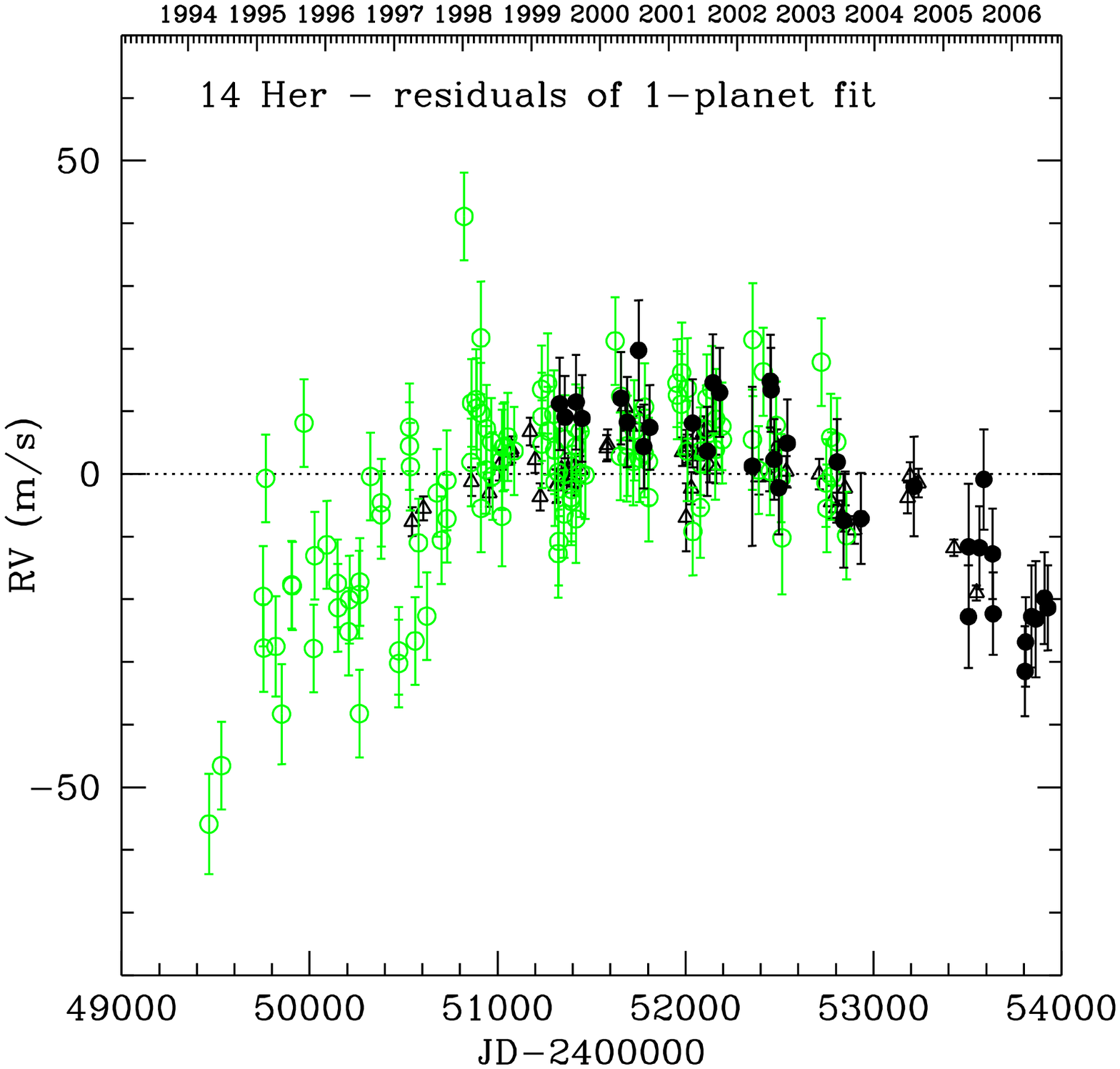}
\caption{Same as Fig.~1, but for a one-planet fit for 14 Her b.  Open 
circles are from 
\citet{naef04}, open triangles are from \citet{butler06}, and filled 
circles are from McDonald. } 
\end{figure}

\begin{figure}
\plottwo{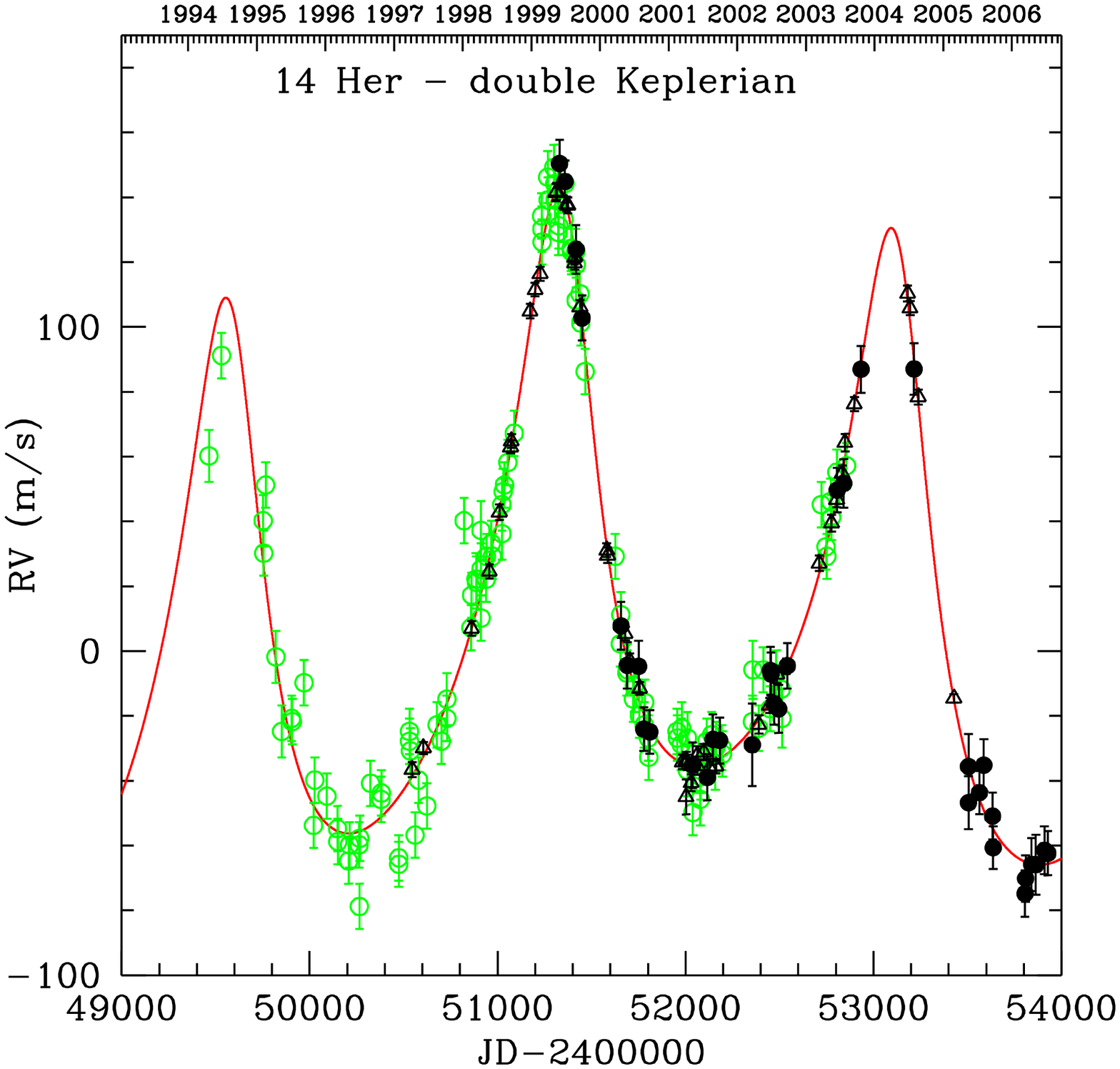}{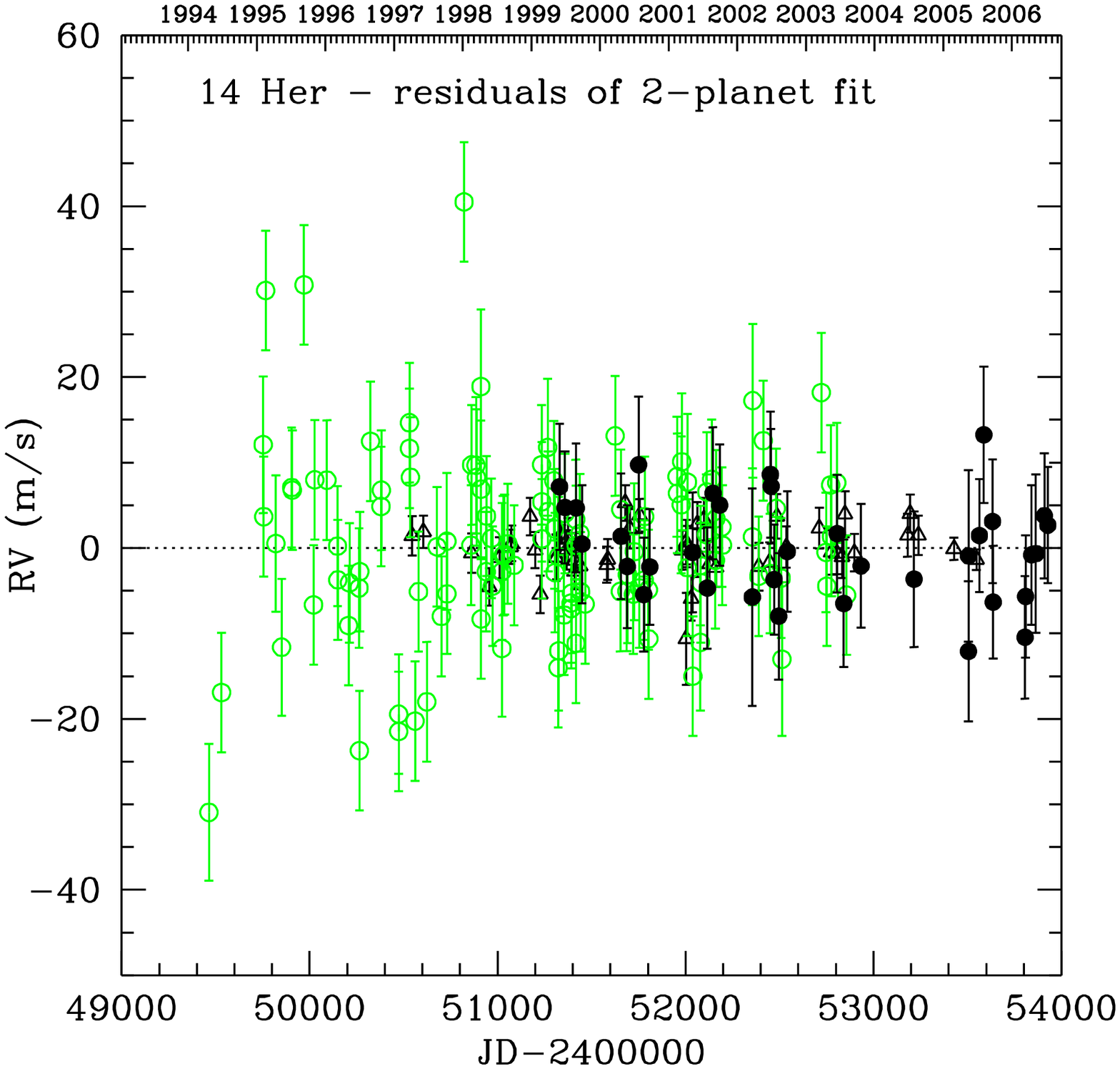}
\caption{Left panel: Double-Keplerian fit for 14~Her.  Right panel: 
Residuals of the 2-planet fit.  Open circles are from
\citet{naef04}, open triangles are from \citet{butler06}, and filled
circles are from McDonald. }
\end{figure}

\end{document}